\pdfoutput=1

\documentclass[epj]{svjour}
\usepackage{graphics}
\usepackage{makecell}
\usepackage{siunitx}
\usepackage{cite}
\usepackage{xcolor}
\usepackage{placeins}
\usepackage{rotating}

\begin{document}
\title{Emergence of simple patterns in many-body systems: from macroscopic objects to the atomic nucleus}

\author{R.F. Garcia Ruiz\inst{1,2}
\thanks{\emph{E-mail: rgarciar@mit.edu}}
\and A.R. Vernon \inst{3, 4}
\thanks{\emph{E-mail: adam.vernon@kuleuven.be}}%
}                     % Do not remove


\institute{CERN, CH-1211 Geneva 23, Switzerland\and Massachusetts Institute of Technology, Cambridge, MA 02139, USA \and KU Leuven, Instituut voor Kern- en Stralingsfysica, B-3001 Leuven, Belgium \and School of Physics and Astronomy, The University of Manchester, Manchester M13 9PL, United Kingdom}
\date{Received: date / Revised version: date}
\abstract{Strongly correlated many-body systems often display the emergence of simple patterns and regular behaviour of their global properties. Phenomena such as clusterization, collective motion and appearance of shell structures are commonly observed across different size, time, and energy scales in our universe.
Although at the microscopic level their individual parts are described by complex interactions, the collective behaviour of these systems can exhibit strikingly regular patterns.
This contribution provides an overview of the experimental signatures that are commonly used to identify the emergence of shell structures and collective phenomena in distinct physical systems.
Examples in macroscopic systems are presented alongside features observed in atomic nuclei.
The discussion is focused on the experimental trends observed for exotic nuclei in the vicinity of nuclear closed-shells, and the new challenges that recent experiments have posed in our understanding of emergent phenomena in nuclei.}

%\PACS{
%      {PACS-key}{discribing text of that key}   \and
%      {PACS-key}{discribing text of that key}
%     }

\titlerunning{Emergence of simple patterns in many-body systems}
\maketitle

\section{Introduction}
\label{intro}
Our understanding of the universe is intimately related to our description of many-body systems. The knowledge of the fundamental particles and forces of nature is as important as our ability to understand how these building blocks are organized to form complex systems. Remarkably, the emergence of simple and regular patterns are common features observed in strongly correlated many-body systems \cite{Teich2016,Reinhardt2005,Arp2004,Grigorieva2006,Wang2018,Suess1956,Fowler1992,Brack1993,DeHeer1993}. At the microscopic level, the individual parts of different physical systems can be described by fundamentally different interactions, however, their collective behaviour can exhibit similar patterns. These seemingly simple regularities of certain properties of a physical system tends to suggest the existence of underlying symmetries and allows simple models to provide a good description of the observed data \cite{Anderson1972,sch81,Gol87,DeHeer1993,Tal93}. However, the link between these models and their microscopic interactions is an open question in many fields of physics. 

The over the past decade experimental and theoretical developments have allowed an unprecedented connection between reductionist and emergent views of nature.
Advances in many-body methods and the rapid development of computing power have provided new paths towards the ab-initio description of macroscopic phenomena. Theoretical developments are motivated by the ambition of a first principles description of emergent phenomena, yet this reductionist approach is deeply motivated by empirical observations \cite{Gib18,Coleman2017}.
A deeper understanding of the microscopic origin of emergent physical phenomena is achieved through systematic experimental studies confronted with the theoretical descriptions.
This article presents a short overview of experimental signatures that are commonly used to characterize the emergence of phenomena in different physical systems. Various examples of objects from the human size scale down to the femtometer scale are presented. The discussion is centered on the observables that are used to indicate the emergence of phenomena such as shell structures and ``magic'' numbers - integer number of constituents with notably different properties. Albeit not exhaustive, an effort is made to include citations that could be useful to direct interested readers to the relevant literature. 
 
The manuscript is divided in two main parts:
The first part provides a brief description of selected examples that illustrate the emergence of regular patters in macroscopic systems and some of their commonalities and differences with similar patterns observed in the atomic nucleus.
The second part is focused on the experimental signatures used to discuss the emergence of collective phenomena and shell structures in nuclei. Commonly discussed properties such as binding energies, nuclear charge radii, excitation energies and transition probabilities are presented. The discussion is expanded using recent experimental results obtained for the ground-state properties of nuclei in the neighborhood of nuclear shell closures.  Finally, an emphasis is made on the trends and open questions that the new observations pose for our current understanding of nuclear structure in different regions of the nuclear chart. 

\section{Emergence of simple patterns in many-body systems}

\begin{figure*}
\centering
\resizebox{1\textwidth}{!}{\includegraphics{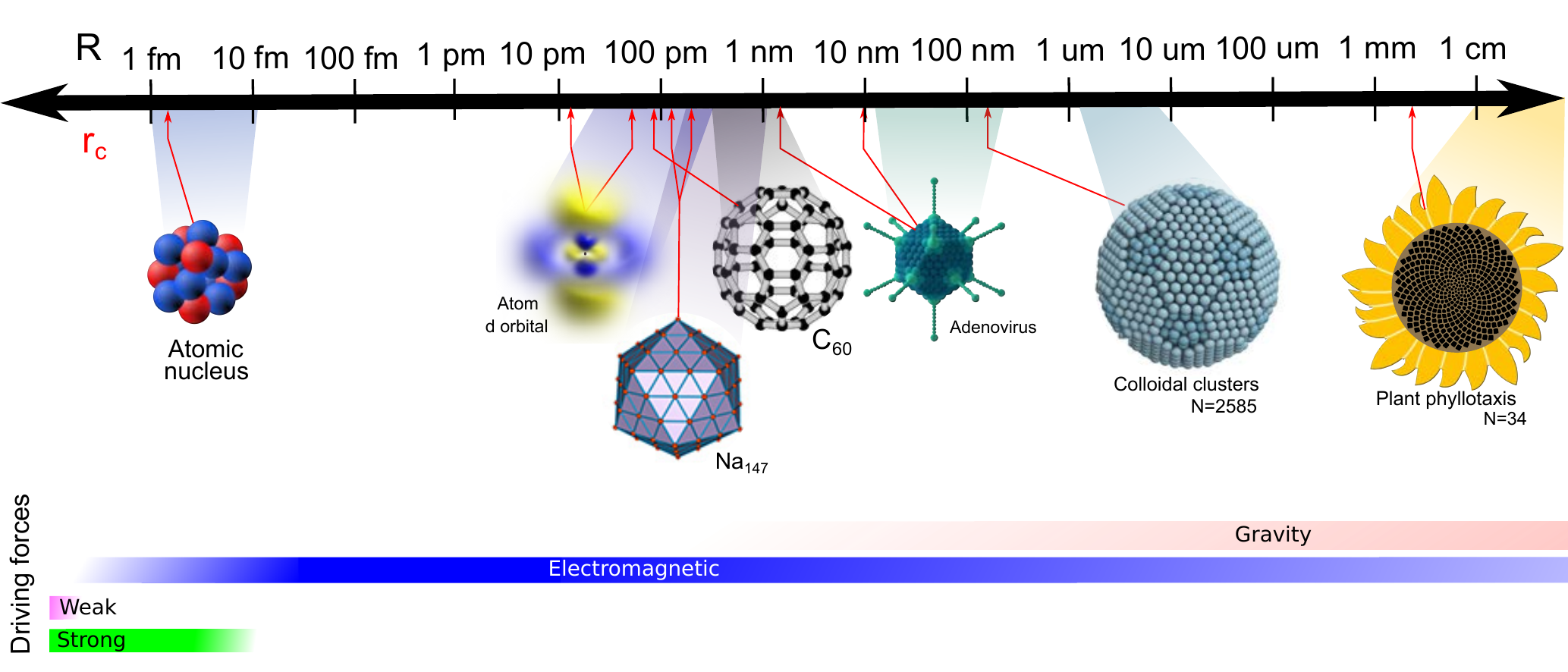}}
\caption{Size scale, R, characteristic length, r$_\text{c}$ and main driving forces are compared for different many-body systems including sunflower seed phylotaxis \cite{NationalMuseumofMathematics}, colloidal clusters \cite{Wang2018}, virus capsid structures \cite{WikimediaCommonsb}, fullerenes \cite{WikimediaCommonsa}, metal clusters \cite{Brack1997}, atoms \cite{WikimediaCommons} and atomic nuclei (Color online).}
\label{fig:systems}
\end{figure*}

Throughout nature, driving forces give rise to optimization problems for the arrangement of constituents in many-body systems at almost every size scale, resulting in an abundance of emergent phenomena \cite{Teich2016, Reinhardt2005, Ridley1982, Vogel1979, Arp2004, Grigorieva2006, Wang2018, Kratschmer1990,Diederich1991,Fowler1992, Echt1981, Echt1987, Knight1984, Pedersen1991, Haberland2005, Sutherland2011,Marenduzzo2010,Ellis2001,Turing1952, Vakili2017}.
On biological scales, despite separation from the smallest constituents of matter by approximately 15 orders of magnitude, simple collective phenomena and pattern formation are still immediately apparent \cite{Teich2016, Cines2014, Hayashi2004a, Kelso1988,Schoner1988}.
Such as in the phyllotaxis of plants \cite{Mitchison1977,Reinhardt2003,Turing1952}, where simple growth patterns appear in the arrangement of leaves or flowers around a plant step.
One particularly striking example is observed in the growth of seeds in a sunflower head \cite{Ridley1982,Vogel1979}, in which the number of spirals of seeds follows the Fibonacci sequence.
A large variety of patterns emerge in smaller systems as a consequence of the optimal arrangement of their constituents \cite{Cines2014, Hayashi2004a, Farhadifar2007}, from the clustering in framboidal pyrite \cite{Ohfuji2005,Ohfuji2002} to the crowding of molecules in cells \cite{Norred2018, Ellis2003, Ellis2001}, or that of DNA strands in cell nuclei \cite{Marenduzzo2010, Micheletti2011}.
Complex many-body systems often form clusters to minimise their energy by interactions between their neighbours and their mean field.
This situation can give rise to ``magic'' numbers, as with those in the atomic nucleus, where certain integer numbers of constituents of a given system results in greater stability of its collective whole.

\begin{table*}
\caption{Experimental signatures of the emergence of shell structures and ``magic'' numbers in different many-body systems.
The size scales and the common observables that are used to characterize the properties of each systems are indicated. Here the ellipses (``...'') are used to denote that additional magic numbers have been omitted for space.}
\label{tab:magic_nos}
\small
\setlength\extrarowheight{0pt}
\begin{tabular}{lp{5mm}llllp{30mm}l}
\hline\noalign{\smallskip}
Constituent & Size & System & Size & $r_c$ & Observable(s) & Magic numbers & Refs. \\
\noalign{\smallskip}\hline\noalign{\smallskip}
Spheres & Any & \makecell[tl]{Spherical \\ packing} & Any & & Density & 6, 12, 21, 25, 38 ... & \cite{Teich2016} \\
Sunflower seeds & $\sim$ 1 cm & \makecell[tl]{Sunflower\\head} & 5-\SI{50}{cm} & \makecell[l]{$\sim$4 mm} & Number of spirals & 3, 5, 8, 13, 21, 34, ... & \cite{Reinhardt2005, Ridley1982,Vogel1979} \\
&\\
Dust particles & \SI{}{\um}  & \makecell[tl]{3D plasma \\crystal} & mm & \makecell[l]{400\\ -\SI{760}{\um}} & Radial distribution & 2, 21, 60, 107 .. & \cite{Arp2004} \\
&\\
\makecell[l]{Superconducting\\ disks} & \SI{}{\um} & Vortex shells & \SI{5}{\um}& $\sim$\SI{2}{\um} & Radial distribution & 5, 16, 32 & \cite{Grigorieva2006} \\
&\\

\makecell[tl]{Polystyrene \\ spheres} & 244 nm & Colloidal cluster & 2-\SI{8}{\um}& \makecell[l]{136 \\-\SI{156}{nm}} & Evaporation rate & 135, 297, 851, 801, 1283, 2583 ... & \cite{Wang2018} \\
&\\
\makecell[l]{GaAs layers} & 10 nm & Quantum dot & \SI{0.5}{\um}& \makecell[l]{7\\ -\SI{10}{nm}} & \makecell[l]{Electron addition \\ energy} & 2, 6, 12, 20 ...  & \cite{Tarucha1996} \\
&\\
\makecell[tl]{Virus protein} & $\sim$5 nm & \makecell[tl]{Virus \\ capsids}  & \makecell[tl]{20\\-\SI{400}{\nm}}& \makecell[l]{2\\ -\SI{10}{nm}} & Abundance & 15,17,18,42 & \cite{Mateu2013,Luque2010,Chen2007,Roos2010, Berger1994} \\
&\\
C atoms & 170 pm & Fullerenes & \makecell[tl]{0.5\\-\SI{2}{nm}} & \makecell[l]{70\\ -\SI{110}{nm}} & Mass abundance & 60, 70, 72, 76, 78, 84 .. & \cite{Goel2004,Kratschmer1990,Diederich1991,Fowler1992} \\
&\\
H$_2$O& 275 pm & \makecell[tl]{Electron-bound \\ water clusters}& $\sim$3-20 \AA& \makecell[l]{170\\ -\SI{190}{pm}}  & Mass abundance & 2, 6, 7, 11 & \cite{Lee2005} \\
&\\
Xe atoms & 216 pm & Atom clusters & $\sim$2-10 \AA & \makecell[tl]{158\\ -\SI{264}{pm}} & \makecell[tl]{Mass abundance} & 13, 16, 19, 25, 55, 71, 87, 147 ... & \cite{Echt1981, Echt1987} \\
&\\
Na atoms & 227 pm & Atom clusters & $\sim$2-10 \AA & & \makecell[tl]{Mass abundance \\ Ionization energy} & 2, 8, 20, 40, 58, 92 ... & \cite{Knight1984, Pedersen1991, Homer1991} \\
 &  &  &  & & Melting temperature & 55, 116, 147, 178 ... & \cite{Haberland2005} \\
&\\
Electrons & fm & Atoms & \makecell[tl]{31\\-348 pm} & \makecell[l]{20\\ -\SI{70}{pm}} & Binding energies & 2, 10, 18, 36, 54 & \cite{A.KramidaY.RalchenkoJ.Reader2014} \\
&\\
Nucleons & fm & Nuclei & 1-10 fm & \makecell[l]{$\sim$ \SI{1.2}{fm}} & \makecell[l]{Binding energies, t$_{1/2}$,} & 2, 8, 20, 28, 50, 126 .. & \cite{Audi2017,Wang2017,Audi2017a, Nordheim1950, Ismail2016} \\
& & & & & $<r^2>$, & & \cite{Angeli2013} \\
& & & & & B(E2), E$_{2^+}$, & & \cite{Pritychenko2016} \\
& & & & & Q$_S$, $\mu$, & & \cite{Stone2016a} \\
& & & & & Solar abundances, & & \cite{Grevesse1998,Suess1956} \\
& & & & & Neutron capture $\sigma$s & & \cite{Ismail2016,Hurwitz1951} \\
&\\
\noalign{\smallskip}\hline
\end{tabular}
\end{table*}

The simplest signature for these magic numbers is the greater natural abundance of systems with certain number of constituents \cite{Kratschmer1990,Diederich1991,Fowler1992, Echt1981, Echt1987, Knight1984, Echt1981, Lee2005}.
This is comparable with the abundance distribution of isotopes in the universe \cite{Alpher1950, Anders1989} following nucleosynthesis \cite{Pagel2009}.
A summary of different systems in nature which exhibit greater stability for certain number of constituents is shown in Table \ref{tab:magic_nos}, where typical experimental signatures and system sizes are highlighted. Examples of the size scale and driving forces in different many-body systems are illustrated in Figure \ref{fig:systems},  where the driving forces refer to those essential for the appearance of the magic numbers in these many-body systems. While gravity, for instance, will have only a minor influence.
The commonality between all of these systems which exhibit magic numbers appears to be: i. a uniformity in the type of constituents; ii. a subtle balance between attractive and repulsive effective forces, which are either self-generated or from an external mean field.
For constituents with their packing constrained by symmetric polyhedral shapes, the magic numbers that appear can be determined for a system with constituents of any size using purely symmetries of geometry, and they appear in nature with these numbers when this is the case \cite{Teich2016, Boles2016, Haberland2005, Echt1981, Rutgers1962, Castillo2008}.
Those constituents between magic numbers, can also be said to belong to a `shell', as with the electronic shells of atoms \cite{A.KramidaY.RalchenkoJ.Reader2014} or for nucleons in atomic nuclei \cite{Mayer1949, Haxel1949}.
In some cases this is reflected by the spatial arrangement of the constituents.
For example in `dusty' plasmas \cite{Bonitz2010,Fortov2005}, where charged dust particles (on the micrometer scale) can self assemble into a plasma crystal arrangement with a radial spherical shell distribution of particles \cite{Arp2005, Arp2004}, with the total system on the scale of millimeters.
Such mesoscopic systems are often called `artificial atoms' due to their close resemblence with atomic systems.
The magic numbers listed in Table~\ref{tab:magic_nos} for dusty plasmas, occur for particular experimental conditions.
These experiments have several highly tunable parameters, which can result in different sequences of magic numbers \cite{Arp2005,Arp2005a,Ludwig2005}, highlighting the ability of a magic number sequence to reflect the underlying interactions of a given system.\\
Self arrangement and the appearance of magic numbers has also been observed in 2-dimensional mesoscopic experiments using micrometer-sized superconducting disks \cite{Juan1998, Grigorieva2006}. At the scale of hundreds of micrometers, polystyrene spheres (colloidal particles) with diameters of around 200~nm have been observed to self-assemble into colloidal clusters \cite{Dinsmore2002,DeNijs2015}.
While the interaction has a complicated description including surface chemistry \cite{Kister2016}, capillary forces \cite{Manoharan2003}, entropy maximization \cite{DeNijs2015} and the presence of depletants \cite{Meng2010}.
These clusters were found to form certain magic number configurations during a drying process \cite{Wang2018}, these configurations were found to have higher thermodynamic stability.
These particularly stable configurations of clusters are found to form upon confinement described mainly by a short-range repulsive potential, and weak attractive interactions between colloidal particles \cite{Wang2018}.
Due to the absence of a long-range repulsive force (such as the Coulomb force which limits the size of atomic nuclei) between the colloidal particles, these colloidal systems can range from a few particles \cite{Manoharan2003} to billions of particles (colloidal crystals) \cite{Clark1979, Velev2000}.

Perhaps the systems with the most in common with the atomic nucleus are atomic clusters \cite{Martin1996,Brack1997,DeHeer1993}, an area of physics which has historically benefited from analogies with nuclear models \cite{Cohen1990,Brack1997,Nishioka1990}.
Clusters of atoms were observed to have magic numbers of enhanced stability reflected in their produced mass abundance spectra \cite{Knight1984, Katakuse1985, Echt1981, Echt1987} (see Table~\ref{tab:magic_nos}).
The electronic structure of the constituent atom ultimately dictates the properties of the atomic clusters, however phenomenological models have been developed to provide a good description of the observed magic numbers, similar to the shell model of the atomic nucleus \cite{Ekardt1984, Brack1993}.
A `wine-bottle' shaped potential used to describe these atomic clusters was adapted from the Woods-Saxon potential of nuclear physics \cite{Katakuse1985}. This potential predicted `super shells' to appear as the number of atoms in the clusters approaches $N$~=~1000 \cite{Nishioka1990}, due to higher-order stabilizing effects, analogous to the predicted islands of stability of heavy nuclei \cite{Li2014,Stoyer2006}.
The predicted super-shell magic numbers were soon observed in sodium clusters \cite{Pedersen1991}.
Deformation also has an analogous role in these clusters as in atomic nuclei, where the most stable clusters have spherical deformation and those between shell closures have oblate or prolate deformation \cite{DeHeer1993, Brack1993}.
The Nilsson model of the atomic nucleus \cite{Nilsson1955} has been adapted to describe axially deformed clusters, known as the Clemenger-Nilsson model \cite{Clemenger1985}.
Giant dipole resonances of atomic nuclei \cite{Berman1975} also have a counterpart in these cluster systems, in the form of plasma resonance frequencies \cite{Ekardt1985,Raza2015}.
Taking the example of the sodium clusters, many of the observables corroborate the same set of magic numbers \cite{Knight1984, Homer1991,Honea1990,Iniguez1986,Wrigge2002} which are of electronic origin.
A modified set of magic numbers was found in the melting temperatures of the clusters \cite{Schmidt2003} however.
This required an additional interpretation considering the geometric shells of the positions of the atomic nuclei alongside the electronic shells, due to the importance of the positions of the atomic nuclei in the melting process \cite{Haberland2005}.

In the fermionic systems, the Wigner-Seitz radius, r$_\text{c}$, is commonly introduced to characterize the length scale of a system (see Ref. \cite{Maruhn2010} for a detailed discussion of fermionic many-body systems).  It is defined as the radius of a sphere whose volume is equal to the mean volume per constituent, given by r$_\text{c}$=R/n$^{1/3}$ for 3-dimensional systems with $n$ constituents, and r$_\text{c}$=R/n$^{1/2}$ for 2-dimensional systems. The r$_\text{c}$ values for some of the systems discussed here are shown in Table~\ref{tab:magic_nos}.
This value also gives an estimate for the density of the systems, which becomes independent of the system size when saturated.
This characteristic is particularly important in the understanding of the saturation of nuclear matter, as the r$_\text{c}$ value remaining nearly constant in finite nuclei, and it remains with a similar magnitude for systems as large as neutron stars \cite{Maruhn2010}.
Similarly, other many-body systems with rigid constituents exhibit comparatively small variations in their r$_\text{c}$ value with the increase of the system size (see Figure~\ref{fig:systems}).

%using only the ground state atomic radii of atoms

Fermionic systems such as atomic nuclei, atoms, atomic clusters and quantum dots can be constrained by similar symmetries \cite{Gaillard1999}. The Pauli exclusion principle, common to all of these systems, imposes the first magic number, 2. However the sequences of magic numbers strongly depend on the forces and symmetries that describe their constituents. 
In the atomic nucleus, the attractive mean-field is self-generated by the constituent nucleons, rather than a common external potential as for electrons in atoms. The `spin-orbit' force between nucleons \cite{Mayer1949}, produces a large attractive force which increases the binding for nucleons in orbital shells with their spins aligned with their orbital angular momentum. Whereas the spin-orbit force between electrons is significantly weaker and repulsive for aligned spin and orbital angular momentum \cite{Foot2004}.

In the influential paper ``More is different'' \cite{Anderson1972}, P.~W.~Anderson argued that as the number of constituents of a system increases, a ``phase transition'' occurs, where the symmetries of the underlying laws of the system are broken and new symmetries can appear, requiring research into these fundamentally new and different laws of the system on the new hierarchy, as was the case with superconductivity \cite{Bardeen1957}.
With this outlook it was said that the ``constructionist'' approach might be lost, that is the ability to predict the emergent phenomena of a system from fundamental laws of physics. It is accepted that emergent properties in complex systems with large number of constituents can require many-body calculations which are presently computationally infeasible, and thus is impossible to establish a direct connection to the fundamental forces of nature  \cite{Altarelli2005, Gaillard1999}. However, for mesoscopic systems such atoms and nucleus, the reductionist viewpoint that the system is still in fact reducible to these laws, is becoming feasible thanks to the developments in powerful many-body methods and computational power. The interactions among electrons are well understood in terms of their underlying theory of quantum electrodynamics, and atomic properties can now be calculated with high accuracy for different  many-electron systems \cite{Pasteka2017a,Eliav2015,Pachucki2006,Pachucki2004,Korobov2001}. The non-perturbative nature of the nuclear force makes the atomic nuclei exceedingly more complex,  and their description from first principles is an ongoing challenge for nuclear theory \cite{Chang2018,Gezerlis2013,Barnea2015,Epelbaum2009a,Ots19,Dyt20}.

\section{Global trends and simple patterns in nuclei} \label{sec:nucleiphenomena}
The atomic nucleus provides a rich laboratory in the studies of strongly correlated many-body systems.
Due to the high nuclear density and the short- and long-range properties of the nuclear force, nuclei are highly sensitive to two- and higher-order many-body forces.
The non-perturbative character of the strong force requires highly demanding theoretical treatments.
In contrast to other physical systems, three-body forces are essential to describe the properties of nuclei \cite{Hammer2013}.
By varying the numbers of protons and neutrons in the nucleus, inter-nucleon correlations can drive very different collective phenomena \cite{Hey11,Lu2013,GarciaRuiz2016,Freer2018}. Intriguingly, a set of regular patterns appear across the whole nuclear chart \cite{Sim86,Cas93,Sorlin2008,Ang15,Bentley2016,Cas19}. These seemingly simple patterns have motivated numerous phenomenological models since the early days of nuclear physics. Simple model principles such as independent-particle motion \cite{Mayer1950,May50b} and the semi-classical collective motion of nuclei \cite{Boh75} have been very successful in providing a global description of the observed nuclear phenomena.
Below we present a short overview of the experimental signatures of nuclear shell structures and collective phenomena that are commonly discussed in literature.
While similar signatures and correlations are found in several nuclear observables, different patterns can emerge in systems with extreme proton-to-neutron ratios.
This discussion is expanded upon using the evolution of nuclear properties in the neighborhood of the neutron-rich $^{52}$Ca ($Z$~=~20, $N$~=~32), $^{78}$Ni ($Z$~=~28, $N$~=~50), and $^{132}$Sn ($Z$~=~50, $N$~=~82) isotopes as examples, where new theoretical and experimental results have become available in the last few years.

\subsection{Experimental signatures of shell structures}
\begin{figure*}
\centering
\resizebox{0.9\textwidth}{!}{
  \includegraphics{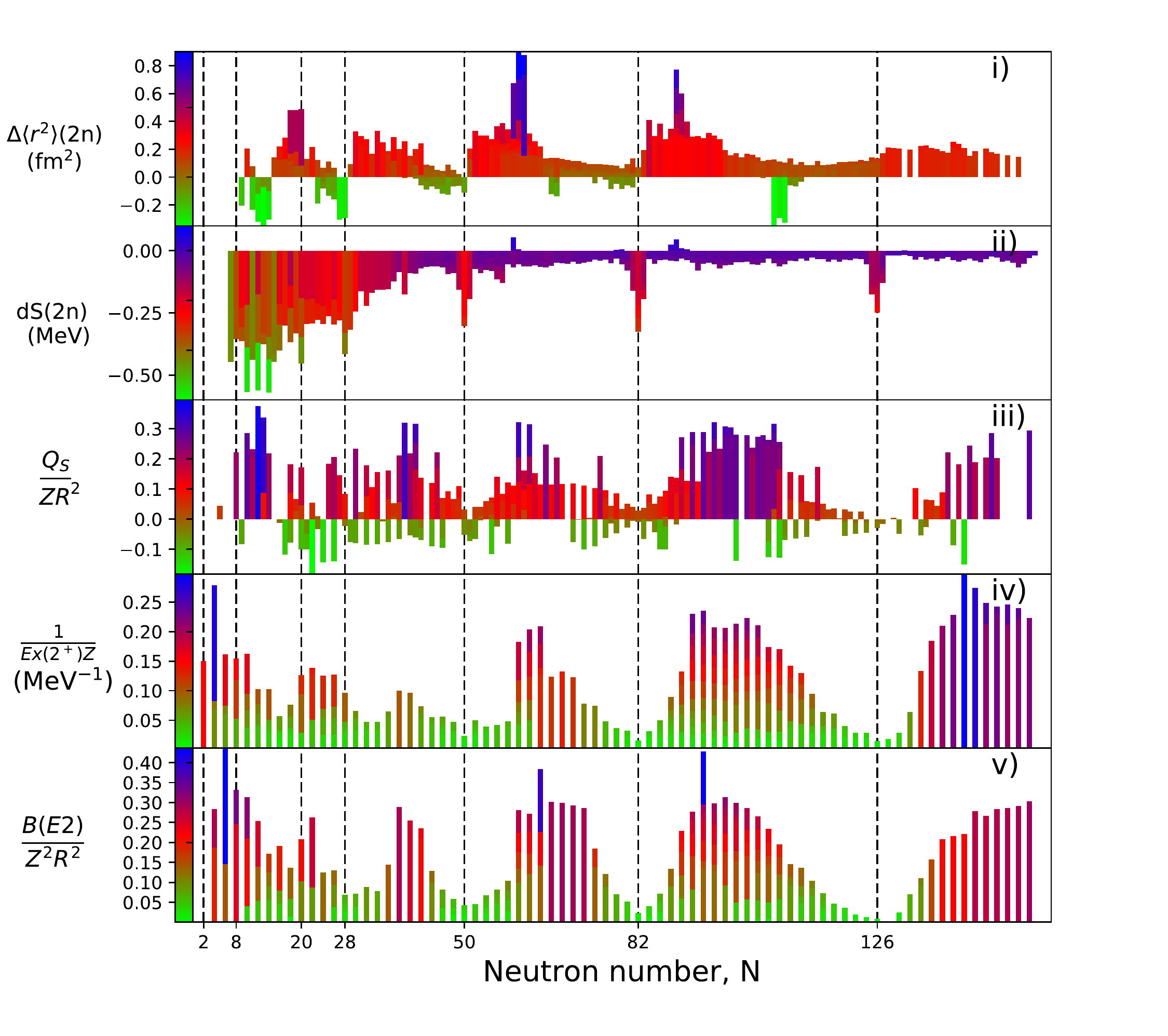}}
\caption{(Color online) Experimental nuclear properties as a function of the neutron number: i. mean-squared charge radii difference when two neutrons are added, $\langle r^2 \rangle$(2n); ii. derivative of the two-neutron separation energy d$S_{2n}$; iii. normalized spectroscopic quadrupole moments $Q_{s}/ZR^2$; iv. scaled inverse of the excitation energy of the first $2^+$ state, $1/E_{2^+}Z$; and v. normalized transition probability to the first $2^+$ excited state, $B(E2)/Z^2R^2$. Data taken from \cite{Audi2017,Wang2017,Audi2017a,Stone2016a,Pritychenko2016,Angeli2013, Eberz1987a,Leb04,Yor13,Gar18,Lindroos1996,
Kreim2014,Bis14,GarciaRuiz2016,Yan16,Han16,Min16,Far17,Hammen2018,Gorges2019,Miller2019a,Xie19,nndc19,GarciaRuiz2015,Babcock2016,Yan18,Mougeot2018,Michimasa2018,Reiter2018a,Klose2019,Liu2019,Xu2019,Leistenschneider2018}. }
\label{fig:prop_N}
\end{figure*}

\begin{figure*}
\centering
\resizebox{0.9\textwidth}{!}{
  \includegraphics{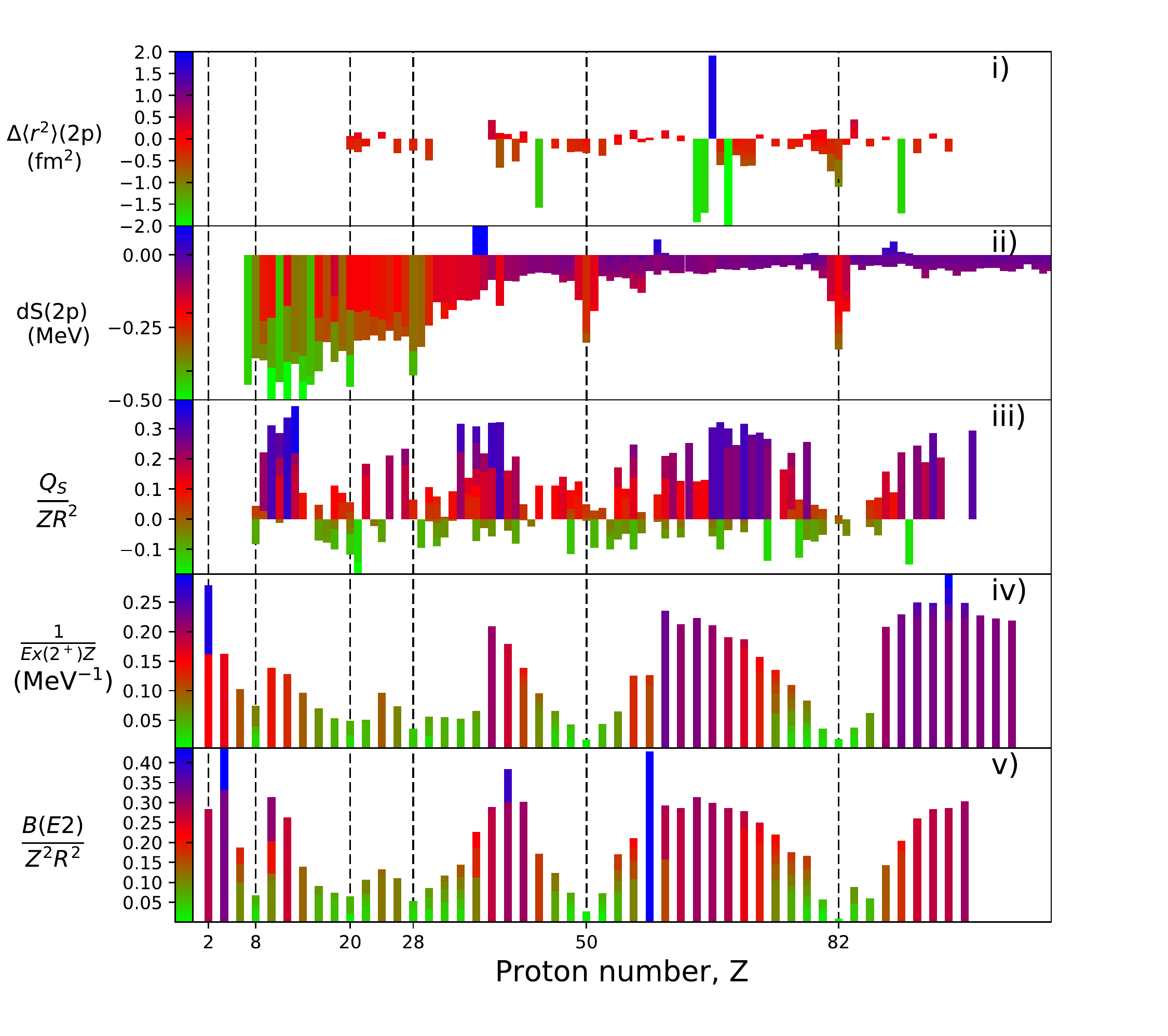}}
\caption{(Color online) Experimental nuclear properties as a function of the proton number: i. mean-squared charge radii difference when two protons are added, $\langle r^2 \rangle$(2p); ii. derivative of the two-proton separation energy d$S_{2p}$; iii. normalized spectroscopic quadrupole moments $Q_{s}/ZR^2$; iv. scaled inverse of the excitation energy of the first $2^+$ state, $1/E_{2^+}Z$; and v. normalized transition probability to the first $2^+$ excited state, $B(E2)/Z^2R^2$.
Data taken from \cite{Audi2017,Wang2017,Audi2017a,Stone2016a,Pritychenko2016,Angeli2013, Eberz1987a,Leb04,Yor13,Gar18,Lindroos1996,
Kreim2014,Bis14,GarciaRuiz2016,Yan16,Han16,Min16,Far17,Hammen2018,Gorges2019,Miller2019a,Xie19,nndc19,GarciaRuiz2015,Babcock2016,Yan18,Mougeot2018,Michimasa2018,Reiter2018a,Klose2019,Liu2019,Xu2019,Leistenschneider2018}.}
\label{fig:prop_P}
\end{figure*}

The signatures of nuclear shell structures are manifested in different observables \cite{Cak10,Sorlin2008,GarciaRuiz2015,GarciaRuiz2016,Wienholtz2013a,Ste13}. The numbers of nucleons that completely fills nuclear closed-shells are the so-called ``magic'' numbers. Nuclei with a magic number of nucleons are commonly observed to have the following experimental signatures:
i. a relatively small mean-squared charge radius, $\langle r^2 \rangle$. As seen in Figure \ref{fig:prop_N} at nucleon number $N=$20, 28, 50, 82 and 126, there is a pronounced change of the charge radius as nucleons are added beyond a shell closure (``kink''), with a smooth increase towards
shell closures, and a larger increase through the filling of the new open shell \cite{Ang15}. ii. a relatively large two-nucleon separation energy, $S_{2n}$;
iii. a small quadrupole moment value, $Q_{s}$;
iv. a high excitation energy of the first $2^+$ state, $E_{2^+}$; and 
v. a small transition probability to the first $2^+$ excited state, $B(E2)$.
A compilation of these experimental properties as a function of the neutron and proton numbers are shown in Figure \ref{fig:prop_N} and Figure \ref{fig:prop_P}, respectively. The data corresponding to different isotones are shown in Figure \ref{fig:prop_N}, using bars of different colors to indicate the magnitudes of the observables for each isotone, the same is shown in \ref{fig:prop_P} as a function of atomic number.

The changes of the mean square charge radii when two neutrons are added, $\Delta \langle r^2 \rangle$(2n), are presented in Figure~\ref{fig:prop_N}. The analogous differences when two protons are added, $\Delta \langle r^2 \rangle$(2p), are shown in Figure~\ref{fig:prop_P}, however the data in this case is relatively sparse as the charge radii of many elements have not yet been measured.
At magic number of nucleons these differences exhibit a minimum value, with local maxima occuring after crossing the closed shell. As the magnitude of $Q_{s}$, $B(E2)$, and $E_{2^+}$ scales with the atomic number and the nuclear size, these parameters were normalized in order to compare light and heavy nuclei on the same scale. The experimental values of $Q_{s}$ and $B(E2)$ were scaled to the dimensionless values $Q_{s}/ZR^2$ and $B(E2)/Z^2R^2$, with $Z$ the proton number and $R=1.18 A^{1/3}$ the droplet-model radius. Normalized observables present minimum values around the nucleon numbers 28, 50, 82 and 126, with a clear correlation seen in the trends of all observables.
For some isotopes, additional local minima appear around nuclear numbers 2, 8, 16, 20, and 40. %This can be seen with bars of different colors.
Figure \ref{fig:prop_N} iv), for example, shows bars of different color at $N$~=~20, indicating that nuclei with the same number of neutrons, such as $^{32}$Mg and $^{40}$Ca, have very different $E(2^+)$ values \cite{Mot95}. 
(We refer the interested reader to Figures~\ref{fig:2dBE2s}-\ref{fig:2dneutseprad} in the appendix for a 2d representation of the data shown in Figures~\ref{fig:prop_N} and \ref{fig:prop_P}.)
The isotopes with magic nucleon numbers have relatively high binding energy, and their charge distribution exhibit smaller variations with respect to the spherical shapes (small quadrupole moments). The nuclear charge radius commonly increases with the number of nucleons, but the slope of the increase is notably smaller approaching the nuclear closed shells. These nuclei are more difficult to excite than their neighbors, which is evidenced by their relatively high excitation energies and low excitation probabilities.

The properties of light nuclei ($A <20$) are highly sensitive to adding or removing a few nucleons. The lower nuclear orbitals have a smaller degeneracy, thus orbital changes can occur for a few nucleons only. 
Some particular isotopes, as in the region around $Z$~=~40, $N$~=~60 and $Z$~=~62, $N$~=~90, are considered to present a rapid onset of deformation \cite{Tog16,Hey11}.
Interestingly, collective phenomena such as shape coexistence and phase transitions observed for nuclei in the region $Z$~=~62, $N$~=~90 have been suggested to exhibit analogous features as those for clusters of silicon atoms, which are governed by very different interactions \cite{Hor06}. 

\subsection{Simple patterns in complex nuclei}
Nuclear electromagnetic moments such as the magnetic dipole and electric quadrupole moment provide complementary insights into the microscopic and collective properties of nuclei \cite{Neyens2003a,Woo13}.
In fact, electromagnetic moments played a key role in motivating the most basic models of nuclear physics: the nuclear shell model \cite{Mayer1950}, and nuclear deformation \cite{Mot76,Bohr1976a,Rai76}. 
Systematic experimental studies of isotopes around nuclear closed shells have revealed surprisingly simple trends in the evolution of nuclear ground-state electromagnetic properties as a function of the neutron number \cite{Eberz1987a,Neyens2003a,Leb04,Yor13,Pap13,Pap14,GarciaRuiz2015,Gro17}.

\begin{figure}
\resizebox{0.5\textwidth}{!}{
  \includegraphics{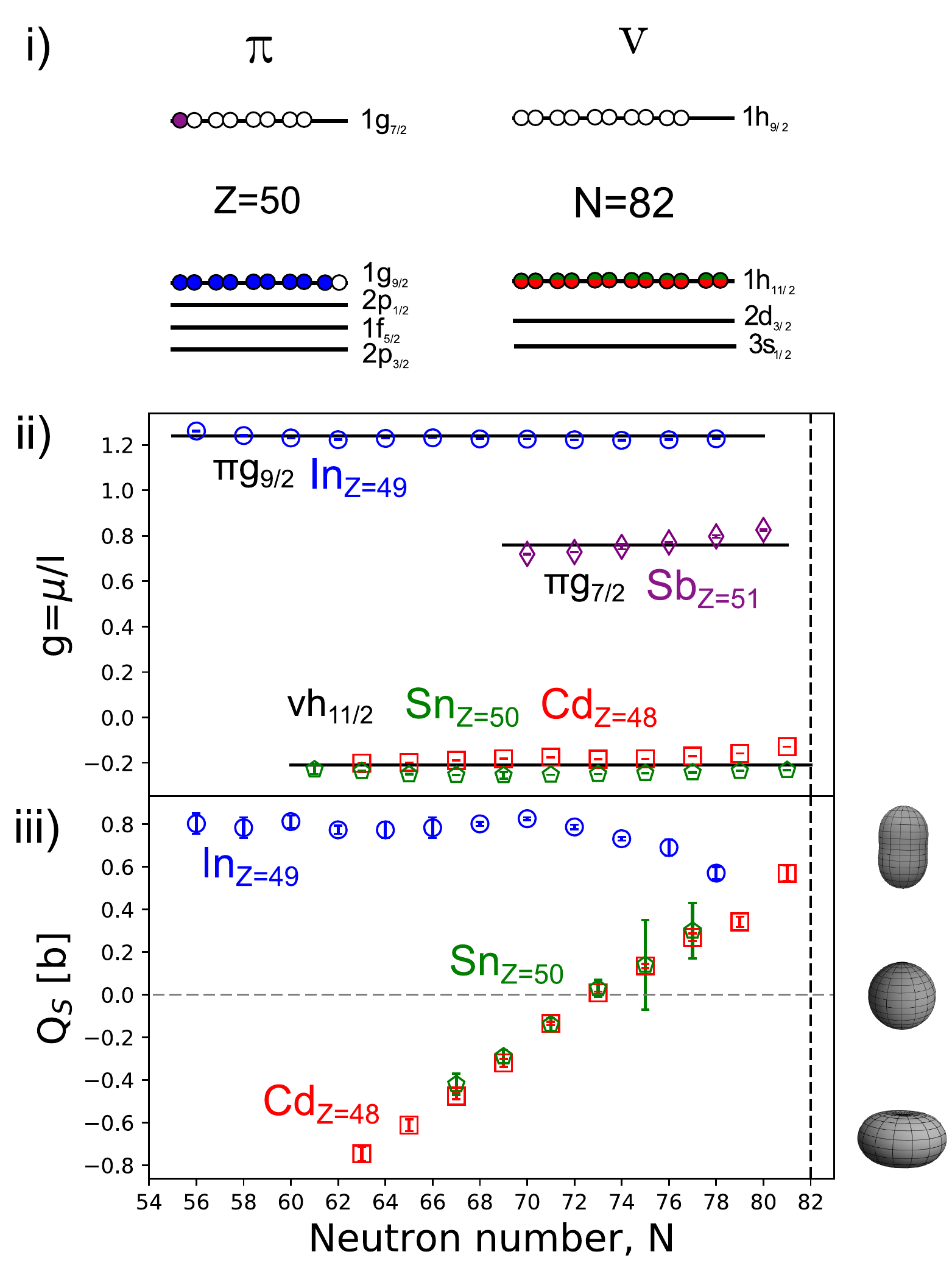}}
\caption{(Color online) Experimental i) nuclear $g$-factor and ii) nuclear quadrupole moments of cadmium ($Z=48$), indium ($Z=49$), and tin ($Z=50$) isotopes. The relevant nuclear shell model orbits are shown on the upper panel. Experimental data were taken from \cite{Eberz1987a,Leb04,Yor13,Gar18,Lindroos1996,Stone2016a}
\label{fig:moments}.}
\end{figure}

Nuclei in the vicinity of the tin isotopes give outstanding examples of simple patterns. The electromagnetic properties of these complex nuclei, with around 50 protons and more than 50 neutrons, seem to be described by a single particle in a nuclear orbital.
The experimental nuclear $g$-factor (the ratio between the dipole magnetic moment and the nuclear spin) and electric quadrupole moments of cadmium ($Z$~=~48), indium ($Z$~=~49), and tin ($Z$~=~50) isotopes are shown in Figure \ref{fig:moments}, exhibiting simple trends as a function of neutron number.
A simplified single-particle model provides a good description of these observations.
In the shell model picture, the electromagnetic properties of odd-even indium isotopes are given by a single proton hole in the $\pi h_{11/2}$ orbit \cite{Arima1954a,Horie1955,Tal63}.
This simple picture of nuclear structure seems to be supported by a rather constant value of their nuclear moments, which present very small variations when neutrons are added.
For the even-proton nuclei, cadmium and tin, the naive shell-model expectation is that that the electromagnetic properties of even-odd isotopes are dominated by a single neutron occupying the $\nu h_{11/2}$ neutron orbit.
This idea is also supported by a constant value of the magnetic moment, and a linear trend in the nuclear quadrupole moments.
In this shell model picture,
a particle occupying an orbit around closed shells has a negative quadrupole moment, which is interpreted as polarizing a spherical core towards an oblate deformation ($Q_{s}<0$) \cite{Neyens2003a}.
If neutrons are added to the same orbit, the values of quadrupole moments cross zero when the orbit is half-filled, and take positive values when more than half of the orbit is occupied. This is interpreted as a ``hole'' polarizing the core towards prolate deformation ($Q_{s} >0$). Similar trends have been observed in the calcium ($Z$~=~20) \cite{GarciaRuiz2015}, nickel ($Z$~=~28) \cite{Wraith2017} and lead ($Z$~=~82) \cite{Neyens2003a} regions.
Although these trends can be interpreted with phenomenological models, the microscopic origin of these remarkably simple emergent trends are not yet explained from first principles.

\begin{figure*}
\resizebox{1\textwidth}{!}{
  \includegraphics{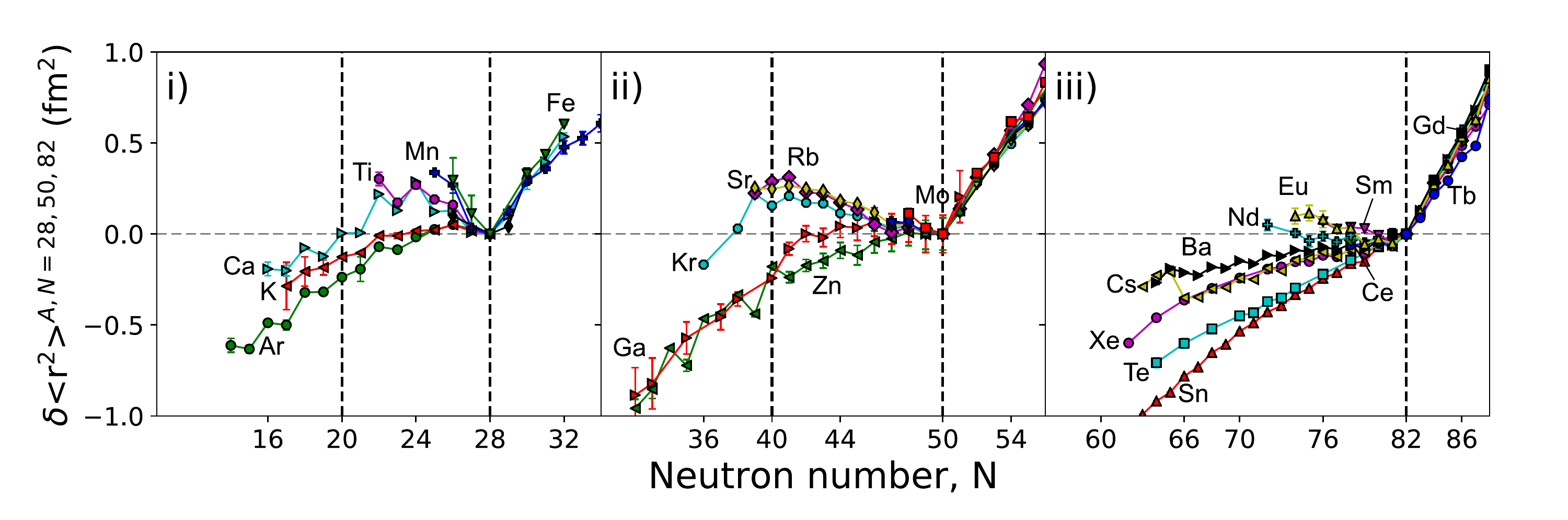}}
\caption{(Color online) Changes in the mean-square charge radii as a function of the neutron number: i. calcium ($Z=20$), ii, nickel ($Z=28$), and iii. tin region ($Z=50$). Each isotopic chain is shown with respect to the isotope with neutron number at the closed shell. Experimental data were taken from \cite{Kreim2014,Bis14,GarciaRuiz2016,Yan16,Han16,Min16,Far17,Hammen2018,Gorges2019,Miller2019a,Xie19}.}
\label{fig:radii}
\end{figure*}

\begin{figure}
\resizebox{0.48\textwidth}{!}{
  \includegraphics{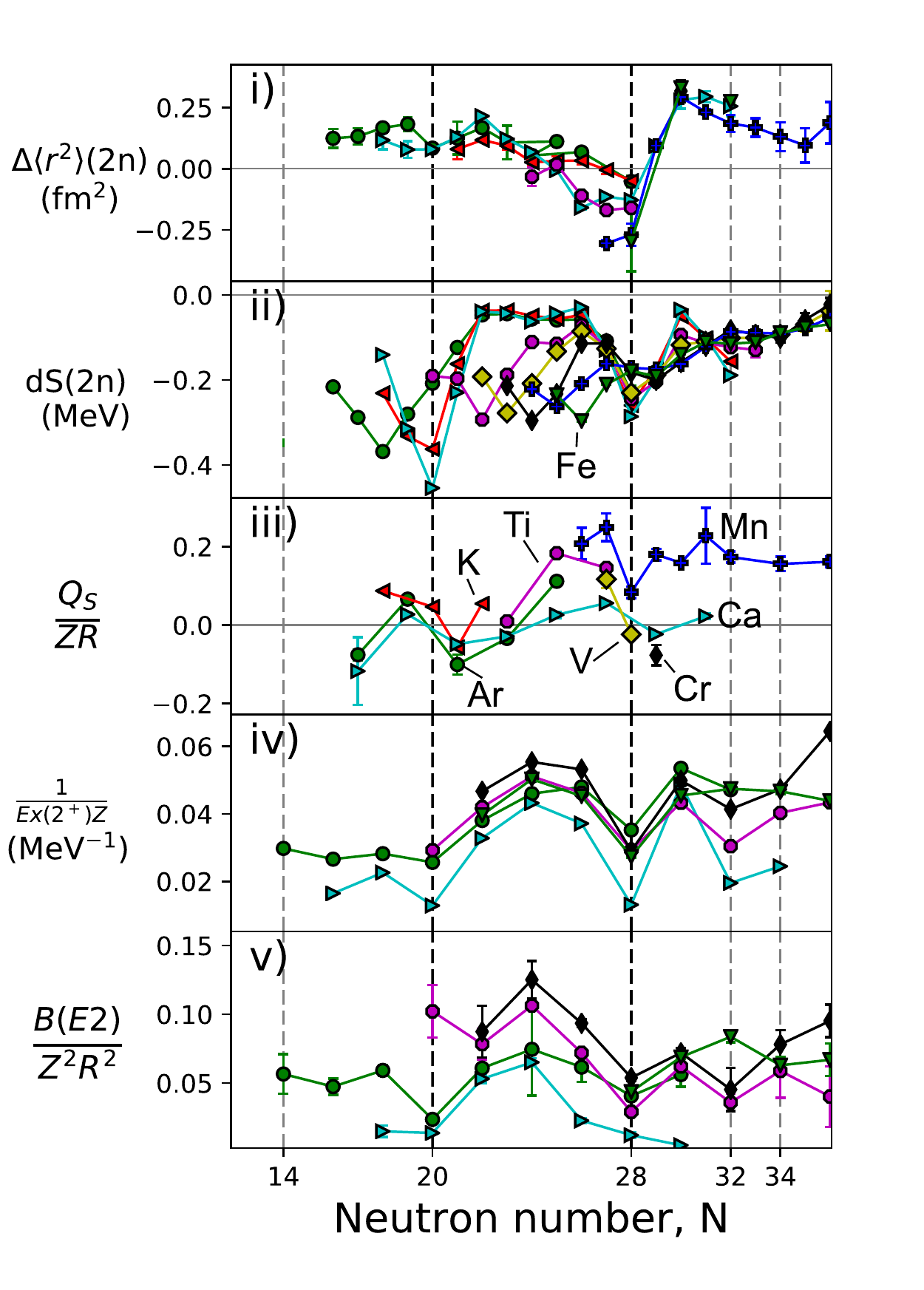}}
\caption{(Color online) Experimental nuclear properties in the calcium region: i. mean-squared charge radii difference when two neutrons are added, $\langle r^2 \rangle$(2n); ii. derivative of the two-nucleon separation energy d$S_{2n}$; iii. normalized spectroscopic quadrupole moments $Q_{s}/ZR^2$; iv. scaled inverse of the excitation energy of the first $2^+$ state, $1/E_{2^+}Z$; and v. normalized transition probability to the first $2^+$ excited state, $B(E2)/Z^2R^2$\label{fig:ca_region}. Experimental data were taken from \cite{Stone2016a,nndc19,GarciaRuiz2016,GarciaRuiz2015,Babcock2016,Yan18,Mougeot2018,Michimasa2018,Reiter2018a,Klose2019,Liu2019,Xu2019,Leistenschneider2018}.}
\end{figure}

\subsection{Selected examples for neutron-rich nuclei}

Recent developments in both experimental and theoretical tools have provided a deeper insight in our understanding of nuclear properties at extreme proton-to-neutron ratios. Particular interest has been focused on the evolution of nuclear properties towards the suggested neutron-rich doubly magic nuclei: $^{52,54}$Ca ($Z$~=~20, $N$~=~32,34)\cite{Kreim2014,GarciaRuiz2016,Wienholtz2013a,Ste13}, $^{78}$Ni ($Z$~=~28, $N$~=~50)\cite{Yan16,Han16,Bis16,Tan19}, and $^{132}$Sn ($Z$~=~50, $N$~=~82) \cite{Gorges2019,Hammen2018}. These regions of the nuclear chart are being studied by several experimental techniques providing tests of theoretical descriptions at limits of the nuclear existence. While most of the measured experimental properties ($S_{2n}$, $E(2^+)$, $B(E2)$, and $Q_s$) have been described by available nuclear models \cite{Wienholtz2013a,Ste13,GarciaRuiz2015,Togashi2018,Tan19}, the description of the nuclear size ($\langle r^2\rangle$) has posed new challenges for modern nuclear theory \cite{Ekstrom2015,GarciaRuiz2016,Lap16,Gorges2019,Miller2019a}.
This problem has been tackled with density functional theory, where satisfactory description of charge radii have been obtained in the calcium \cite{Miller2019a} and tin regions \cite{Gorges2019}. However, a description in the ab-initio framework has not been achieved yet \cite{GarciaRuiz2016,Lap16}. 

Figure \ref{fig:radii} shows the changes of the mean-squared charge radii around the calcium, nickel, and tin regions. The values for each isotopic chain are shown with respect to the value at the closed neutron shells. While a strong element dependence is seen close to stability, the charge radii of neutron-rich isotopes beyond the neutron closed-shell appear to increase with surprisingly similar slopes. The radii of the proton-closed-shell calcium isotopes increase as rapid as the open shell isotopes Mn ($Z$~=~25) and Fe ($Z$~=~26). 

Similar charge radii trends have been observed for isotopes around the nickel and tin regions. As illustrated in Figure \ref{fig:radii} ii), the nuclear charge radii evolution in the nickel region present a noticeable dependence with the atomic number up to the neutron number $N$~=~50. However, for neutron-rich nuclei the mean-squared charge radii of different elements increase with the same slope. Beyond $N$~=~50, the radii of isotopes near to the proton closed-shell such as zinc ($Z$~=~30) increase with the same magnitude as the open proton shell isotopes krypton ($Z$~=~36) and rubidium ($Z$~=~37). These trends are almost identical in the tin region below and beyond the neutron number $N$~=~82 (see Figure \ref{fig:radii}iii ).

The rapid increase of the nuclear charge radii observed beyond the neutron number $N$~=~28 is in contrast with the patterns seen in isotopes close to stability.
For neutron-rich nuclei in the calcium region, the discontinuities seen in other observables such as $S_{2n}$ \cite{Wienholtz2013a} and $E(2^+)$ \cite{Ste13} values at neutron number $N$~=~32, do not appear to be evident in the nuclear charge radii trends. A compilation of different properties measured in the calcium region is shown in Figure \ref{fig:ca_region}.  The signatures of closed shells at $N$~=~20 and $N$~=~28 appears across all observables. For the nuclear charge radii (Figure \ref{fig:ca_region}i ) the signatures at $N$~=~20 are present but less pronounced than for $N$~=~28. At $N$~=~32 and $N$~=~34 the clear agreement for the signs of shell closures among the different observables breaks down, and distinct regular patterns appear for different observables. 

Only very recently systematic measurements have been achieved for the nuclear charge radii in the vicinity of calcium and tin isotopes beyond $N$~=~28 and $N$~=~82 \cite{GarciaRuiz2016,Han16,Gorges2019}.
The charge radii and electromagnetic moments of $^{58-70}$Ni, $^{124-134}$Sn and $^{112-134}$Sb isotopes have been measured by the COLLAPS collaboration at ISOLDE-CERN \cite{Ett19,Gorges2019}. Moreover, results for $^{47-52}$K ($Z$~=~19), $^{58-78}$Cu ($Z$~=~29), $^{104-111}$Sn ($Z$~=~50) and $^{101-131}$In ($Z=49$) isotopes have been obtained by the CRIS collaboration at ISOLDE-CERN \cite{Kos19,Kos19b,DeGroote2019,Ver19}.
Efforts are underway to extend these measurements to more exotic neutron-rich isotopes beyond the $^{52}$Ca, $^{132}$Sn nuclei \cite{GarciaRuiz2016,Gar17,Ruiz2017b,Kos19b}
Excitation energy measurements of the doubly-magic shell closure of $^{78}$Ni indicates the onset of significant structural changes in neutron-rich isotopes beyond this region \cite{Tan19}.
Measurements of the nuclear charge radii of isotopes around $Z$~=~28, $N$~=~50 and beyond is a challenging area for present experimental studies \cite{DeGroote2019}.
The development of radioactive beam facilities \cite{Thoennessen2014} and experimental techniques will be required to allow for nuclear charge radii measurements of lighter neutron-rich isotopes in the oxygen region.
Their measurement will give insight into the onset of this seemingly nuclear size independent gradient of charge radius increase with neutron number following a shell closure. \\

\section{Conclusions}
Despite the drastic difference in the interactions between their constituents, the collective properties of strongly correlated many-body systems exhibit  common features. From dust particles governed by Coulomb interactions, atomic clusters interacting by covalent bonds and inter-atomic potentials, up to nuclei governed by short-range nuclear forces.
The interactions, length scale and dynamics are very different, but these systems present similar signatures of shell structures and collective phenomena.
The commonalities between these many-body systems have shown to be fruitful to allow for mutual advancements in different fields, as for example was found in the field of atomic nanoclusters by the successful application of modified nuclear structure models.
The recent developments in many-body theory and the continuous increase in computing power have allowed an unprecedented reductionist insight of the emergence of physical phenomena.
In complex correlated many-body systems, where accurate calculations are particularly challenging, the connection between reductionist and emergence viewpoints is commonly guided through empirical observations. 
In contrast to other quantum systems, the atomic nucleus is formed by two different constituents (protons and neutrons) that interact mainly by the electromagnetic, strong and weak forces. Moreover, three-body forces appear at a fundamental level in the strong interaction \cite{Hammer2013a}.

Recent developments in many-body methods and higher computing power have provided great steps towards the understanding of the microscopic origin of collective phenomena in different regions of the nuclear chart \cite{Lau16,Hag16E,Ots19,Caurier2005,Tog16,Leo17}. 
However, forming a consistent and unified microscopic description of the distinct nuclear phenomena remains as an open problem for nuclear theory. 
A particular challenge has been the description of nuclear charge radii, where new data in neutron-rich nuclei all exhibit an intriguingly simple increase in charge radii beyond nuclear closed-shells. Moreover, the electromagnetic properties of isotopes around magic numbers of protons and neutrons have been found to exhibit astonishingly simple trends. The microscopic description of these simple patterns, which are predicted by the oldest models of nuclear physics, is a major challenge for modern nuclear theory.

\section*{Acknowledgements}
This work was supported by ERC Consolidator Grant No.648381 (FNPMLS); STFC grants ST/L005794/1,\\ ST/L005786/1, ST/P004423/1 and 
Ernest Rutherford Grant No. ST/L002868/1; 
GOA 15/010 from KU Leuven,
BriX Research Program No. P7/12;
the FWO-Vlaanderen (Belgium); 
the European Unions Grant Agreement 654002 (ENSAR2).
We thank A. Koszorus and S. Wilkins for helpful comments and suggestions.

\onecolumn

\section*{Appendix}

\FloatBarrier

\renewcommand{\thefigure}{A\arabic{figure}}
\setcounter{figure}{0}

\begin{figure*}[!h]
\resizebox{0.9\textwidth}{!}{
  \includegraphics{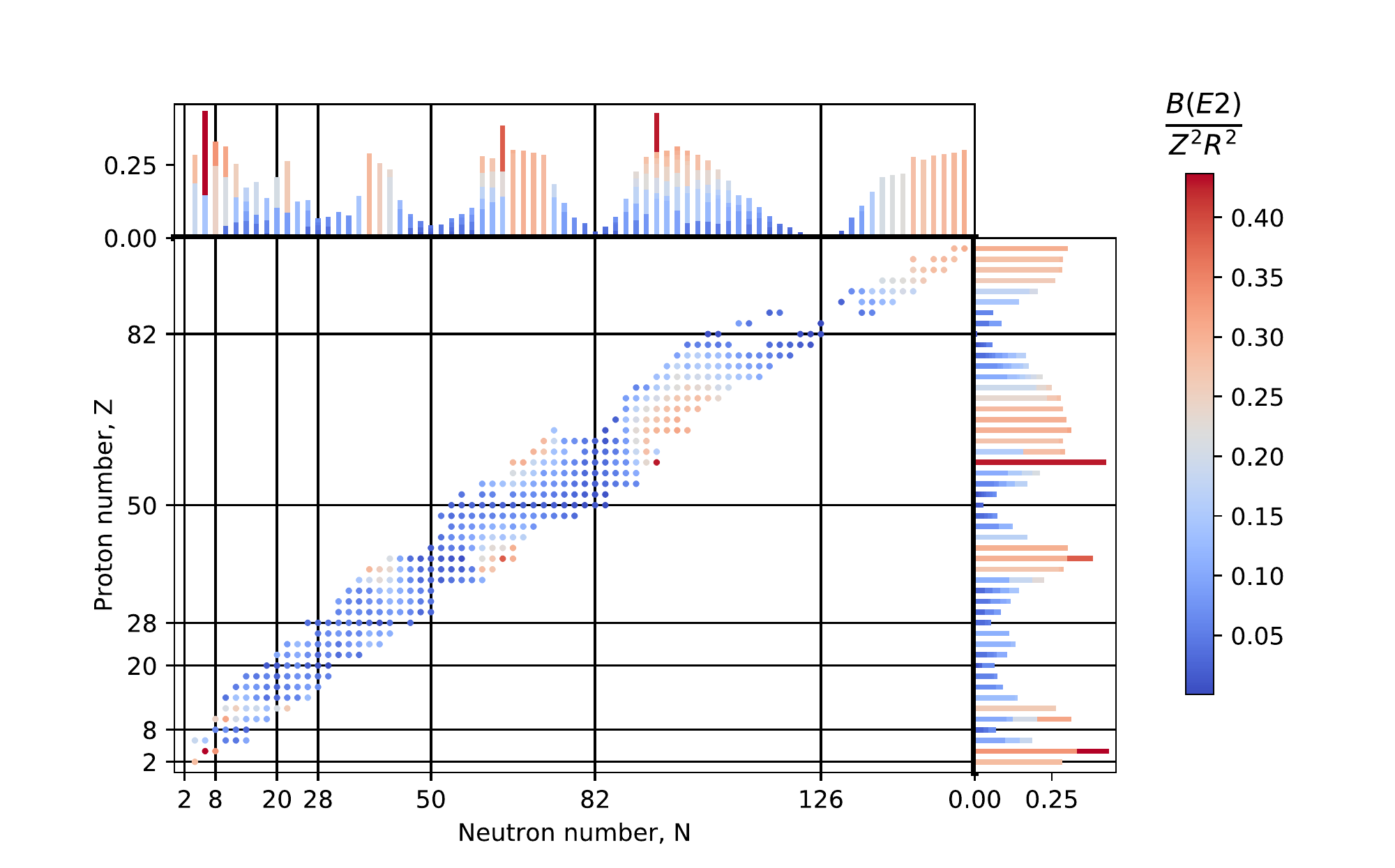}}
\caption{(Color online) Normalized transition probability to the first $2^+$ excited state, $B(E2)/Z^2R^2$, as a function of the proton number and neutron number.
Data taken from \cite{Audi2017,Wang2017,Audi2017a,Stone2016a,Pritychenko2016,Angeli2013, Eberz1987a,Leb04,Yor13,Gar18,Lindroos1996,
Kreim2014,Bis14,GarciaRuiz2016,Yan16,Han16,Min16,Far17,Hammen2018,Gorges2019,Miller2019a,Xie19,nndc19,GarciaRuiz2015,Babcock2016,Yan18,Mougeot2018,Michimasa2018,Reiter2018a,Klose2019,Liu2019,Xu2019,Leistenschneider2018}.}
\label{fig:2dBE2s}
\end{figure*}

\begin{figure*}[!h]
\resizebox{0.9\textwidth}{!}{
  \includegraphics{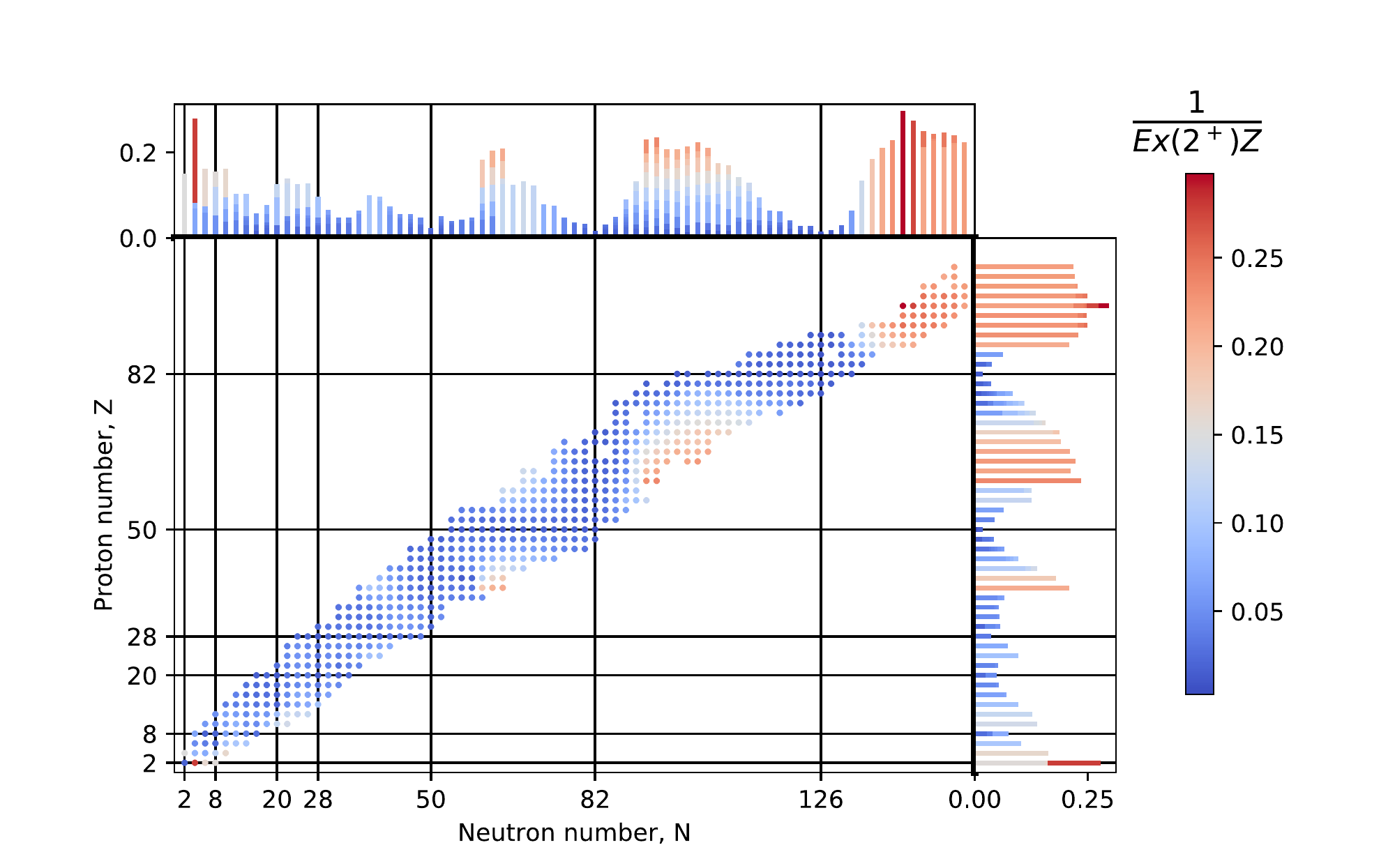}}
\caption{(Color online) Scaled inverse of the excitation energy of the first $2^+$ state, $1/E_{2^+}Z$, as a function of the proton number and neutron number.
Data taken from \cite{Audi2017,Wang2017,Audi2017a,Stone2016a,Pritychenko2016,Angeli2013, Eberz1987a,Leb04,Yor13,Gar18,Lindroos1996,
Kreim2014,Bis14,GarciaRuiz2016,Yan16,Han16,Min16,Far17,Hammen2018,Gorges2019,Miller2019a,Xie19,nndc19,GarciaRuiz2015,Babcock2016,Yan18,Mougeot2018,Michimasa2018,Reiter2018a,Klose2019,Liu2019,Xu2019,Leistenschneider2018}.}
\label{fig:2dEx2s}
\end{figure*}

\begin{figure*}[!h]
\resizebox{0.9\textwidth}{!}{
  \includegraphics{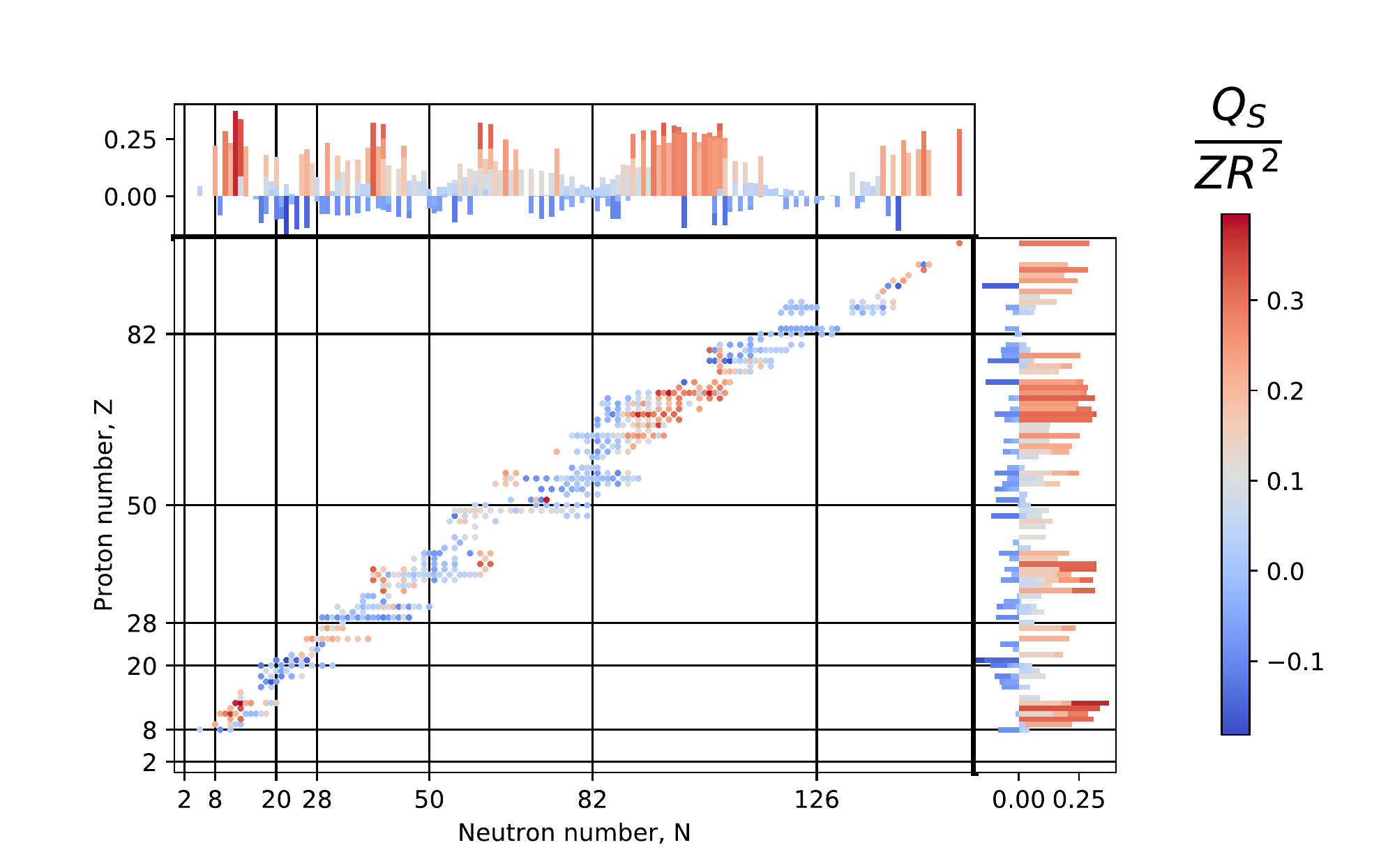}}
\caption{(Color online) Normalized spectroscopic quadrupole moments $Q_{s}/ZR^2$, as a function of the proton number and neutron number.
Data taken from \cite{Audi2017,Wang2017,Audi2017a,Stone2016a,Pritychenko2016,Angeli2013, Eberz1987a,Leb04,Yor13,Gar18,Lindroos1996,
Kreim2014,Bis14,GarciaRuiz2016,Yan16,Han16,Min16,Far17,Hammen2018,Gorges2019,Miller2019a,Xie19,nndc19,GarciaRuiz2015,Babcock2016,Yan18,Mougeot2018,Michimasa2018,Reiter2018a,Klose2019,Liu2019,Xu2019,Leistenschneider2018}.}
\label{fig:2dQs}
\end{figure*}

\begin{figure*}[!h]
\resizebox{0.9\textwidth}{!}{
  \includegraphics{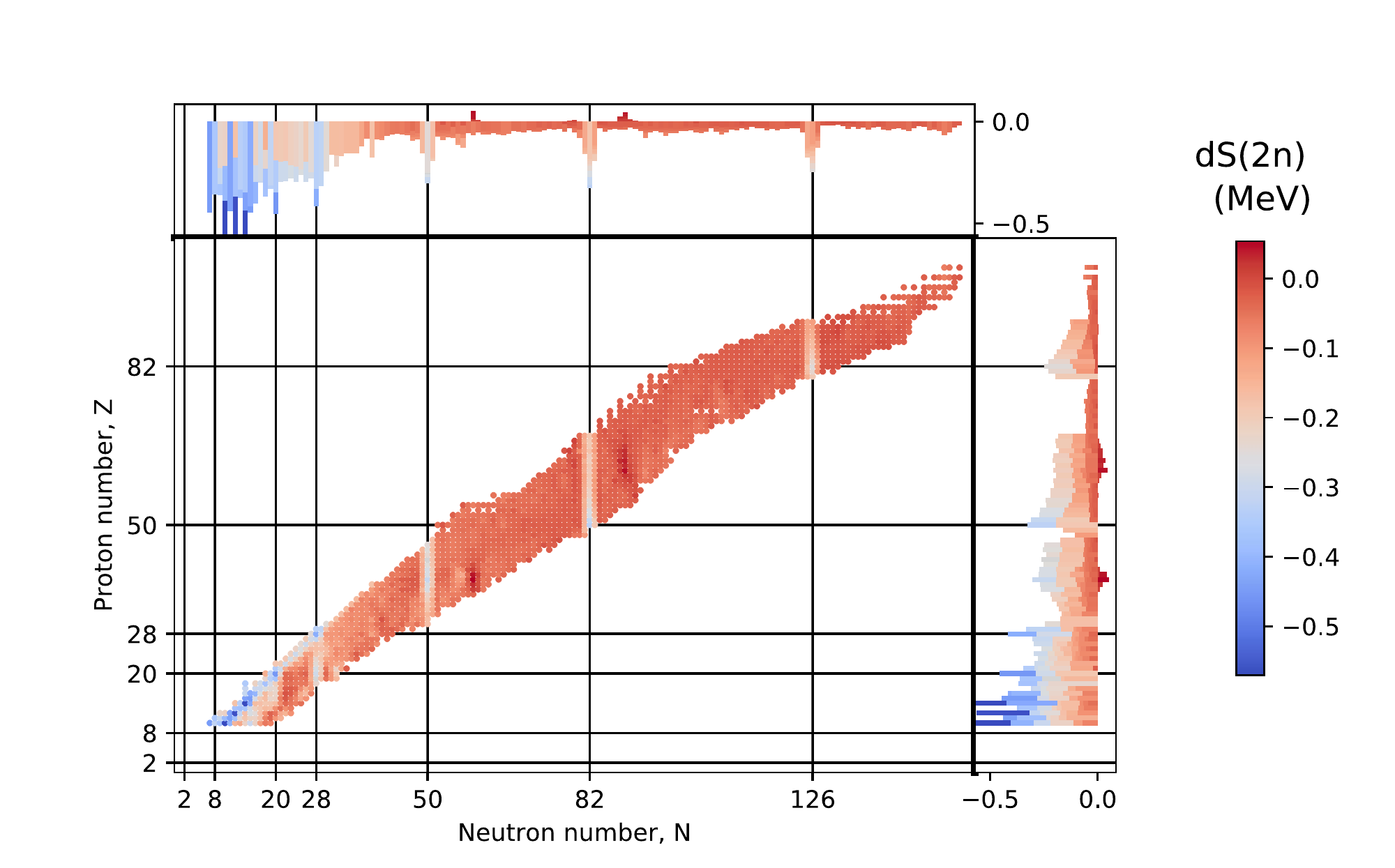}}
\caption{(Color online) Derivative of the two-proton separation energy d$S_{2n}$, as a function of the proton number and neutron number.
Data taken from \cite{Audi2017,Wang2017,Audi2017a,Stone2016a,Pritychenko2016,Angeli2013, Eberz1987a,Leb04,Yor13,Gar18,Lindroos1996,
Kreim2014,Bis14,GarciaRuiz2016,Yan16,Han16,Min16,Far17,Hammen2018,Gorges2019,Miller2019a,Xie19,nndc19,GarciaRuiz2015,Babcock2016,Yan18,Mougeot2018,Michimasa2018,Reiter2018a,Klose2019,Liu2019,Xu2019,Leistenschneider2018}.}
\label{fig:2dneutsepenergy}
\end{figure*}

\begin{figure*}[!htb]
\resizebox{0.9\textwidth}{!}{
  \includegraphics{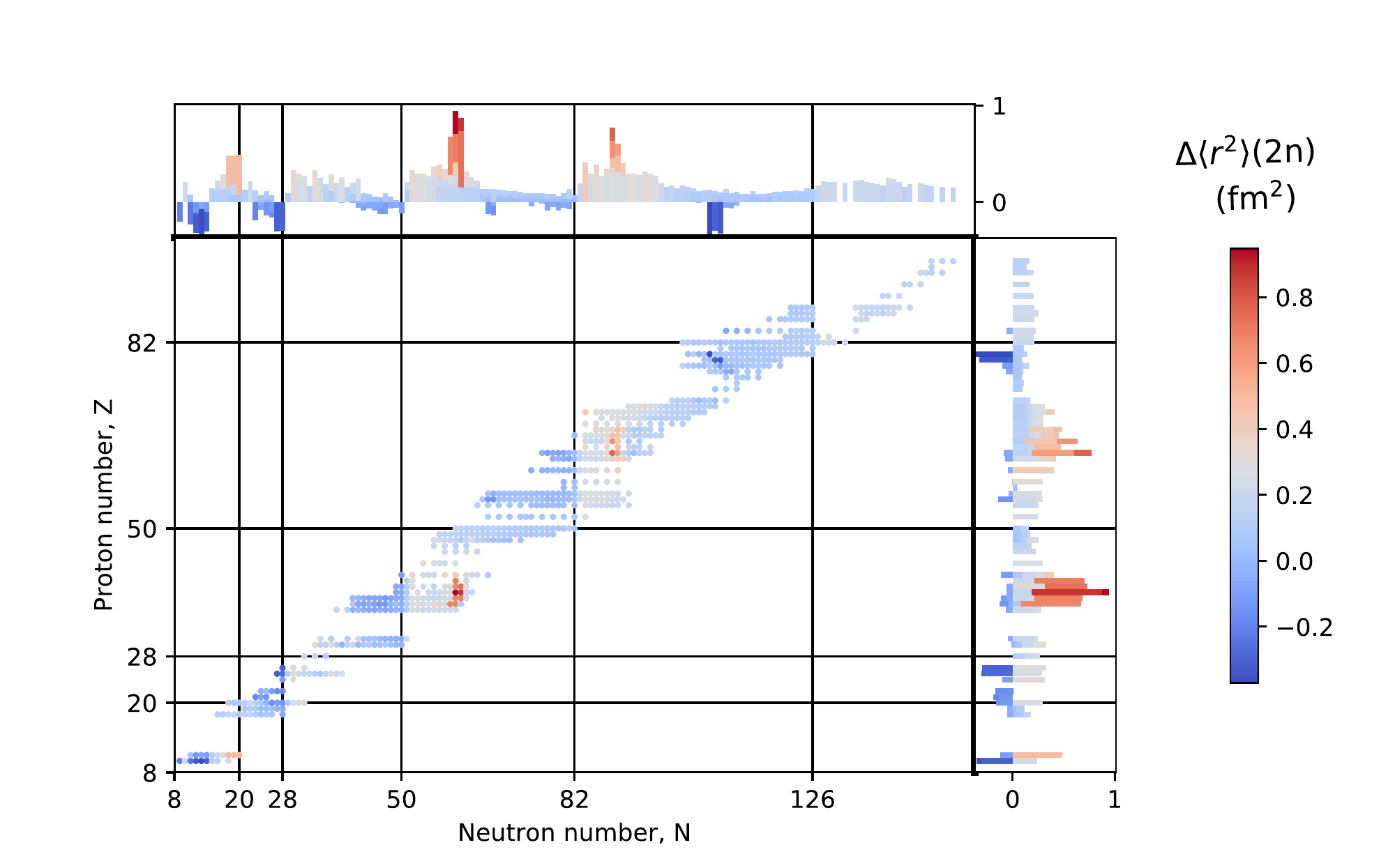}}
\caption{(Color online) Mean-squared charge radii difference when two protons are added, $\langle r^2 \rangle$(2n), as a function of the proton number and neutron number.
Data taken from \cite{Audi2017,Wang2017,Audi2017a,Stone2016a,Pritychenko2016,Angeli2013, Eberz1987a,Leb04,Yor13,Gar18,Lindroos1996,
Kreim2014,Bis14,GarciaRuiz2016,Yan16,Han16,Min16,Far17,Hammen2018,Gorges2019,Miller2019a,Xie19,nndc19,GarciaRuiz2015,Babcock2016,Yan18,Mougeot2018,Michimasa2018,Reiter2018a,Klose2019,Liu2019,Xu2019,Leistenschneider2018}.}
\label{fig:2dneutseprad}
\end{figure*}

\clearpage

\FloatBarrier

\twocolumn

\bibliographystyle{unsrt}
\bibliography{main}

\end{document}